\documentclass[english,aps,twocolumn]{revtex4}
\usepackage[T1]{fontenc}
\usepackage[latin1]{inputenc}
\usepackage{amsmath}
\usepackage{amssymb}

\makeatletter
\usepackage{bm}

\usepackage{graphics}

\usepackage{epsfig}

\usepackage{color}

\usepackage[ps2pdf,colorlinks=false, pagebackref=false, bookmarks=true,bookmarksopen=true,bookmarksnumbered=true]{hyperref}

\newcommand{\Cu}{\ensuremath{\mathrm{Cu}}}

\newcommand{\Al}{\ensuremath{\mathrm{Al}}}

\makeatother

\usepackage{babel}
\makeatother
\begin{document}
\def\Cu{Cu$_{2}$MnAl }

\def\Al{AlO$_x$ }

\title{Evidence for triplet superconductivity in Josephson junctions with ferromagnetic Cu$_{2}$MnAl-Heusler barriers}

\author{D. Sprungmann$^{*}$ , K. Westerholt$^{*}$ , H. Zabel$^{*}$ , M. Weides$^{**}$ ,and  H. Kohlstedt$^{***}$}

\affiliation{$^{*}$Institut f\"ur
Experimentalphysik\:/\:Festk\"orperphysik, Ruhr-Universit\"at
Bochum, 44780 Bochum, Germany; $^{**}$Physics Department, Broida
Hall, UC Santa Barbara, CA 93106, USA; $^{***}$Nanoelektronik, Technische Fakultät Kiel, 
Christian-Albrechts-Universität Kiel, Kiel 24143, Germany}

\begin{abstract}
We have studied Josephson junctions with barriers prepared from
the Heusler compound Cu$_2$MnAl. In the as-prepared state the \Cu layers
are non ferromagnetic and the critical Josephson current density
$j_{c}$ decreases exponentially with the thickness of the Heusler
layers $d_{F}$. On annealing the junctions at 240°C the Heusler
layers develop ferromagnetic order and we observe a dependence
$j_{c}(d_{F}$) with $j_{c}$ strongly enhanced and weakly thickness
dependent in the thickness range 7.0 nm < $d_{F}$ < 10.6 nm. We
attribute this feature to a triplet component in the
superconducting pairing function generated by the specific
magnetization profile inside thin \Cu layers.

\pacs{74.45+c, 74.25.Nf, 74.78.Fk}

\end{abstract}
\maketitle

Superconducting pairing functions with a symmetry different from
conventional s-wave singlet pairing are in the focus of interest
since the advent of BCS theory
\cite{Werthamer}. However, unconventional pairing functions are
rare in nature and experimental realizations had to wait until detecting systems 
like heavy fermions- \cite{UPt2}, high-T$_{c}$´s-\cite{hightc} and the Sr$_2$RuO$_{4}$-superconductors
\cite{Srruthenates}. Unconventional pairing states might also be
induced by the superconducting proximity effect at
superconducting/ferromagnetic (S/F) interfaces (see
\cite{Buzdinrev},\cite{Efetovrev}  for recent reviews). The
exchange field in the ferromagnetic layer favors triplet
pairing i.e. a superconducting condensate function with parallel
spins. The penetration depth of superconducting triplet pairs into
a ferromagnet in the limit of dirty metals with weak pair breaking
scattering is given by $\xi=\sqrt{\hbar D_F/2\pi k_BT}$
(electron diffusion constant $D_{F}$) whereas the penetration
depth of singlet pairs is limited by the ferromagnetic exchange
energy $E_{ex}$ via $\xi_{F}$= $\sqrt{\hbar D_F/E_{ex}}$
\cite{Buzdinrev}. Thus at low temperatures triplet pairs can
penetrate deeply into a ferromagnetic metal.

In early experiments \cite{Girou} the resistance change of a
ferromagnetic nanowire crossing a superconducting nanowire below
the superconducting transition temperature was analyzed and
the results were interpreted in favor of triplet
superconductivity. The best experimental verification of long
range triplet superconductivity would be the direct measurement of
the decay length of the superconducting pairing function, as can
be done by measuring the critical current $j_{c}(d_{F})$ in
Josephson junctions with ferromagnetic barrier layers of variable
thickness $d_{F}$. However, up to now the
experimental studies of $j_{c}(d_{F})$ did not indicate 
long range triplet superconductivity \cite{JJthicknessdependence}.
Singlet pairing with a transition from 0- to $\pi$-coupling  
(0- and $\pi$ denoting the phase shift of the pair wave function
across the barrier) could reasonably well explain all data.

Recent theoretical work suggested an interesting possibility to
enhance the amplitude of the triplet component at the S/F
interface drastically. Instead of a homogeneous magnetization profile as in Ref.\cite{JJthicknessdependence} one should better use a ferromagnetic layer system with some
intrinsic spin canting e.g. an in-plane Néel wall at the interface
\cite{BergeretTriplet}, an S/F/S multilayer with non parallel
orientation of the F-layers \cite{Bergeret2003}, or a trilayer
system with canted spins on both sides facing the S-layers
\cite{HouzetTriplet}. 
The triplet component in the condensate function is enhanced by conversion of singlet Cooper pairs into 
triplet pairs by spin active interfaces \cite{EschrigTriplet}. The dominating superconducting wavefunction is the so
called odd-frequency triplet
pairing, i.e. a superconducting wave function even in space, even
in spin but odd in time (or odd in the Matsubara frequencies).
This exotic type of pairing state has been proposed originally for
the pairing in the He$^{3}$ superfluid \cite{Berezinskii}.

Transport measurements on lateral Josephson junctions
with half metallic CrO$_{2}$ \cite{KeizerTriplet} and the Rare
Earth metal Ho \cite{SosninTriplet} as the barrier material gave
first indications of weak long range triplet contributions to the
supercurrent. In the present Letter we will show that the unique 
properties of thin films of the Heusler alloy Cu$_2$MnAl can create a magnetization profile inside 
the Cu$_2$MnAl layer, which effectively generates triplet superconductivity.

\begin{figure}
\centering
\includegraphics[width=8.0cm]{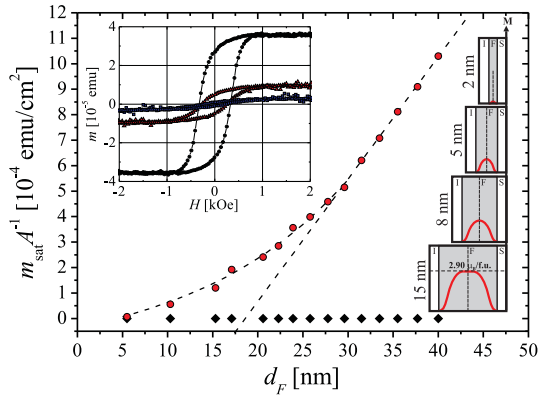}

\caption{(Color online) Magnetic moment (measured at 1 kOe and 15
K) divided by the sample area versus the thickness of the Heusler
layer in the as-prepared state (diamonds) and after annealing at
240°C for 24 h (dots). The dashed linear slope corresponds to a magnetic moment of
2.9 $µ_{B}$/Mn-atom. The left inset shows hysteresis loops measured at
$15$ K for samples in the annealed state with the Heusler layer
thickness $d_F$ = 5.4 nm (inner loop, blue), $d_F$ = 10.2 nm (middle loop, red) and $d_F$ = 17 nm (outer loop, black).
The right inset schematically depicts the magnetization profile within Cu$_2$MnAl-layers of different thicknesses.} 
\label{figure01}
\end{figure}

Our SIFS-Josephson junctions with a lateral size of $10\times50\:{\rm
{\mu m^{2}}}$ or $10\times200\:{\rm {\mu m^{2}}}$ were fabricated
by dc-sputtering and micro-structured by optical lithography and ion beam
etching, as described in detail in Ref. \cite{WeidesPhysicaC}. The
barrier between the two Nb-electrodes is composed of an about 1 nm
thick \Al-layer prepared by thermal oxidation of Al and a
Cu$_2$MnAl-layer with increasing thickness along the wafer axis. The thin \Al layer determines the normal resistance $R_n$ and allows a precise
analysis of the critical current without changing the physics
discussed above. In a single preparation run we prepared several
hundred junctions covering the thickness range from typically 5 nm
to 15 nm for the Heusler alloy while keeping the other layer
thicknesses constant. The I-V -characteristics of the junctions
were measured in a flow cryostat in the temperature range between
1.8 K and 6 K using home made, fully automatized electronics. The
critical current was defined using a voltage drop criterion of
$\sim$0.5 \ensuremath{µ}V. Transport measurements were done on Josephson junctions 
in the as-prepared state as well as after annealing
at 240°C for 24 h in an UHV oven. During this annealing process,
which does not degrade the
junction, the Heusler layers become ferromagnetic. The magnetic
properties of the thin Heusler layers were studied by a SQUID magnetometer on similar but
not micro-structured layer stacks.

\begin{figure}
\centering
\includegraphics[width=8.0cm]{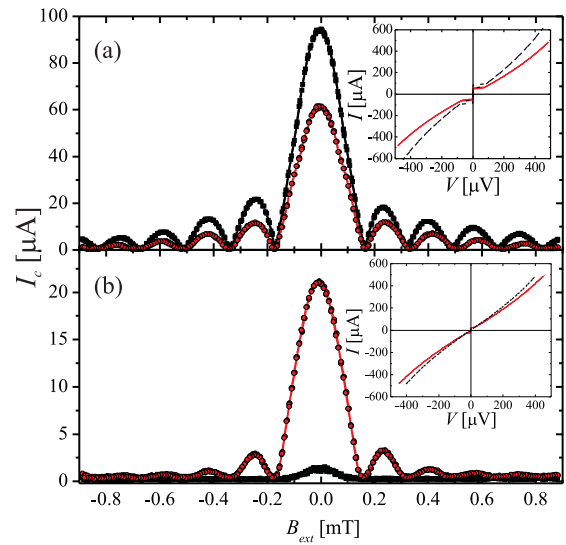}
\caption{(Color online) Critical current versus the applied
magnetic field for two junctions ($10\times50\:\rm{\mu m^2}$) with a thickness of the Heusler
layer $d_F$ = 6.7 nm  (a) and $d_F$ = 9.3 nm (b) in the as-prepared state
(black squares) and the annealed state (red circles). 
Insets: I-V characteristics measured at zero magnetic field. Solid curves: annealed state; dashed curves: as-prepared state.
All measurements done at $T=4.2$ K.} 
\label{figure02}
\end{figure}

The magnetization measurements of our samples with the Cu$_2$MnAl-layers
in the as-prepared state reveal that they are non-ferromagnetic
(see Fig.\ref{figure01}). When prepared at room temperature \Cu grows in the
A2-structure i.e. with a random distribution of all atoms on a bcc
lattice \cite{Heuslerreview}. Nearest neighbor
Mn-atoms are coupled by strong antiferromagnetic exchange
interactions whereas Mn-next nearest neighbors with an Al-atom
in-between are coupled ferromagnetically via a RKKY type
interaction \cite{Kuebler}. The competition of these two
interactions and the randomness of the atomic distribution in the
A2 structure gives rise to a spin glass type magnetic order as seen in Fig.\ref{figure01}. Upon
annealing at 240°C the unit cell symmetry transforms to the
ordered L2$_{\text{1}}$-type Heusler structure which is combined
of four interpenetrating fcc-sublattices occupied by Mn, Cu and
Al, exclusively \cite{Heuslerreview}. With this symmetry in \Cu
there are no Mn-Mn nearest neighbors and the ferromagnetic
Mn-Al-Mn exchange interactions lead to ferromagnetic order with a
magnetic moment of about 3.2 \ensuremath{µ}$_{\text{B}}$ per
Mn-atom and a ferromagnetic Curie temperature of 603 K
\cite{Heuslerreview}.

In the left inset of Fig.\ref{figure01} we show examples of ferromagnetic hysteresis
loops for different thicknesses of the Cu$_2$MnAl-layers
after annealing at 240°C for 24 h. A ferromagnetic hysteresis loop
is observed above a critical thickness $d_{F}$ = 5 nm. For $d_{F}$
< 5 nm the spin glass state still exists. Above $d_{F}$ = 5 nm the
saturation magnetic moment increases gradually over a very broad
thickness range up to about $d_{F}$= 30 nm before it reaches a
constant slope corresponding to a value of 2.9
\ensuremath{µ}$_{\text{B}}$/Mn-atom, similar to the bulk
saturation moment. This behavior of the Heusler layers is
very unique and indicates a gradual transition of the magnetic
order from pure spin glass to high moment ferromagnetism across
intermediate phases with coexisting spin glass order and low
moment ferromagnetism. The magnetization profile (right inset in Fig.\ref{figure01}) inside each
Heusler layer also reflects these different phases. We expect low
moment spin glass type of order close to the interfaces and larger
moment ferromagnetic type of order in the core of the film. The
microscopic origin of this behavior is an intrinsic gradient of
the degree of L2$_{\text{1}}$-type atomic order inside the Heusler
layer with a low degree of order at the interfaces and a higher
degree of order in the interior of the film \cite{GrabisBergmann}.

\begin{figure}
\centering
\includegraphics[width=8.0cm]{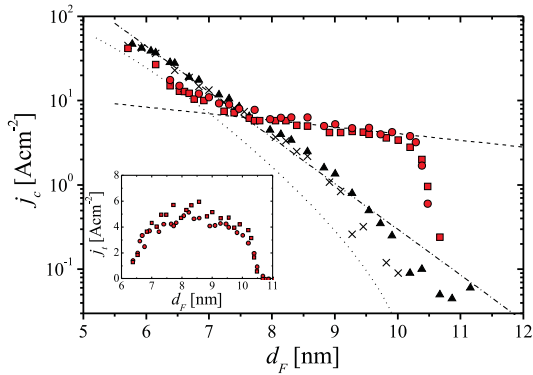}
\caption{(Color online) Critical current density (log) versus the Heusler layer thickness in the
as-prepared state (crosses: $10\times50\:\rm{\mu m^2}$ and triangles: $10\times200\:\rm{\mu m^2}$) and the annealed state (circles: $10\times50\:\rm{\mu m^2}$ and squares: $10\times200\:\rm{\mu m^2}$). 
Measurement done at 4.2 K. Dashed line: $\xi=5.7$ nm; dash-dotted line: $\xi=0.8$ nm; dotted line: singlet component model $j_s(d_F)$. The thick solid line denotes the instrumental resolution limit. 
Inset: Triplet supercurrent $j_t(d_F)$ calculated by subtracting $j_s(d_F)$ from $j_c(d_F)$ (see main text).}
\label{figure03}
\end{figure}

The I-V-characteristics of a Josephson junction with a Heusler barrier
thickness $d_F$ = 6.7 nm before and after annealing is shown in Fig.\ref{figure02}. At $T=4.2$ K we observe
the characteristics of an underdamped junction up to a thickness
range of $d_F$ = 7.5 nm in the as-prepared state and 9.5 nm after
annealing. Beyond this thickness range the Josephson dynamics is overdamped, as typical for junctions with thicker
ferromagnetic barriers due to the high subgap currents \cite{WeidesPRLAPL}. For all 
junctions we observe text book-like Fraunhofer patterns (Fig.\ref{figure02}) with the
critical current vanishing at the minima giving clear evidence of
a high quality and homogeneity of both barriers (\Al and Cu$_2$MnAl).

The thickness dependence of the critical current density in the
as-prepared state is plotted in Fig.\ref{figure03}. One observes an
exponentially damped curve with a decay length $\xi$ = 0.8 nm, as
typical for a system with strong pair breaking scattering. Strong
inelastic pair breaking scattering in the Heusler layers must be
expected, since in the spin glass state there is a high density of
randomly oriented Mn magnetic moments. The decay length of a dirty
metal ($\ell_m\ll \xi_F$) with inelastic pair breaking scattering ($E_{ie}\gg k_BT$) is given by
$\xi_N=\sqrt{\hbar D_F/E_{ie}}$ with the diffusion constant $D_F$ and the
scattering energy $E_{ie}=\hbar/\tau_{ie}$ (scattering time
$\tau_{ie}).$ Single Cu$_2$MnAl layers in the as-prepared state have a residual resistivity of 
$\rho_{m}\approx275\:{\rm {\mu\Omega cm}}$
and with the Fermi velocity $v_F$ taken from the literature
\cite{Fenander1968} we estimate $D_F$ and get $E_{ie}$ = 45 meV
for the inelastic scattering energy.

\begin{figure}
\centering
\includegraphics[width=8.0cm]{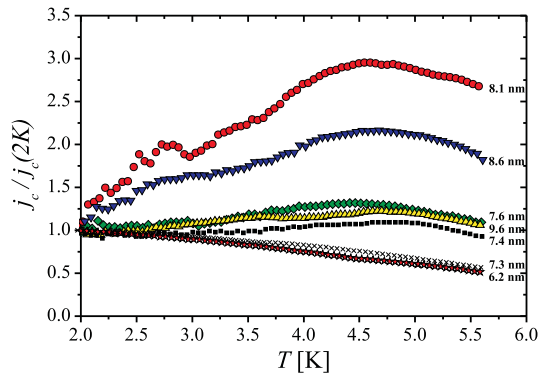}
\caption{(Color online) Normalized critical current density versus
temperature for junctions with different Heusler thicknesses (as-prepared: $d_F$ = 7.3 nm, annealed: all others, $A=10\times50\:\rm{\mu m^2}$).} 
\label{figure04}
\end{figure}

In Fig.\ref{figure02} we show characteristic examples of Fraunhofer
patterns for two junctions before and after annealing. The sample with $d_F$= 6.7 nm has a slight decrease of the critical
current after annealing, as typical for the whole thickness range
below $d_F$ = 7 nm (see Fig.\ref{figure03}). This decrease is due to a
slight increase of the $AlO_{x}$ barrier resistance $R_{n}$. From the observed
change of the residual electrical resistivity upon annealing we
find that $D_{F}$ increases by about a factor of $\sim 3$.

For the second sample with the thickness $d_F$ = 9.3 nm in Fig.\ref{figure02},
the critical current in the annealed state is a factor of 14
larger than in the as-prepared state! This is very surprising,
since the increasing ferromagnetic exchange field should suppress
rather than enhance the critical current carried by singlet Cooper pairs. As plotted in Fig.\ref{figure03},
this enhancement is observed for all samples in the thickness
range between 7.5 nm and 10.5 nm, the enhancement factor reaches a
maximum value of nearly 100 close to $d_{F}$ = 10.5 nm. Above 10.5
nm there is a sharp drop-off and the critical current approaches
the value for the as-prepared state again. In order to verify our data we prepared two sample series in this 
thickness range with equal AlO$_x$ barriers. Both of them are included in Fig.\ref{figure03} and follow the same curve, including
the sharp drop-off at $d_{F}$ = 10.5 nm.

The theory for singlet pairing of junctions with a ferromagnetic
barrier of variable thickness $d_{F}$ in the dirty limit predicts
the functional dependence
\vspace{-0.1cm}
\begin{eqnarray}
%j_{c}(d_{F})=j_{0}\left|\cos{\left(\frac{d_{F}}{\xi_{F2}}\right)}\right|\exp{\left(-\frac{d_{F}}{\xi_{F1}}\right)}\label{fit}
j_{c}(d_{F})=j_{0}\left|\cos{\left(d_{F}/\xi_{F2}\right)}\right|e^{-d_{F}/\xi_{F1}}
\label{fit}
\end{eqnarray}

with the decay length $\xi_{F1}$ and the oscillation length
$\xi_{F2}$ \cite{Buzdinrev}. Without pair breaking scattering both
lengths are given by the penetration depth $\xi_{F}$ \cite{Buzdinrev}. The first
term in equation \ref{fit} describes the Josephson phase transition between 0 and 
$\pi$ and leads to a deviation of the current density
from an exponentially damped curve towards smaller values when
approaching the transition range, in sharp contrast to what we
have observed for the annealed state in Fig.\ref{figure03}. For the decay
length in the plateau region in Fig.\ref{figure03} we derive a value of
$\xi$ = 5.7 nm, which in the framework of singlet
superconductivity cannot be interpreted in any reasonable way. For
singlet superconductivity the critical current density should
decay following the initial slope in Fig.\ref{figure03}., approximately,  i.e.
with a decay length $\xi_{F1}\leq 1$ nm. We thus are led to
the conclusion that in the plateau region of Fig.\ref{figure03} an additional
component of the supercurrent appears, which compensates the
exponentially damped singlet supercurrent. The theoretical work
\cite{Buzdinrev},\cite{Efetovrev} strongly suggests that this
additional component should be identified as an odd-frequency
triplet supercurrent. The superposition of a short range singlet and a long range triplet component ($j_t(d_F)$)
becomes more apparent if we assume a model function for the singlet supercurrent $j_s(d_F)$
corresponding to Eq. \ref{fit} and subtract it from the total supercurrent. 
For the dotted line in Fig.\ref{figure03} we assumed 
an exchange and a scattering energy of $E_{ex}=E_{ie}\approx 60$ meV, which seems   
reasonable, because in many alloys the exchange and the inelastic scattering energy 
have the same magnitude \cite{JJthicknessdependence}. 

As further piece of evidence for the occurrence of unconventional
superconducting pairing in the plateau region of Fig.\ref{figure03}, we show
the temperature dependence of the critical current in Fig.\ref{figure04}. 
For $d_F<7$ nm $j_{c}$ increases with decreasing $T$, this is
the conventional behavior for a singlet supercurrent. Within the
plateau region we observe a broad maximum in $j_{c}(T)$ at about
4.5 K with $j_{c}$ decreasing for lower temperatures. As shown in
the theoretical work \cite{EschrigLTP2007} this anomalous
temperature dependence results from the competition of different
contributions to the supercurrent in junctions with unconventional
pairing. For instance, it can be caused by the superposition of a triplet and a singlet supercurrent 
with opposite Josephson coupling.

Our assumption of the appearance of triplet supercurrent being
responsible for the $j_c$-plateau in Fig.\ref{figure03} implies that inside the
Heusler layers there is a magnetization profile which very
effectively generates triplet superconductivity. Referring to
Fig.\ref{figure01}, one finds that the $j_{c}$-plateau coincides with a narrow
thickness range just above the onset of ferromagnetic order. For
slightly larger thicknesses, but still rather small values of the
saturation magnetization, the extra critical current becomes
rapidly suppressed again. This indicates that the microscopic
origin for the conversion of the singlet into triplet pairs
depends sensitively on the magnetization profile inside the
Heusler layers. In the thickness range just above the onset of
ferromagnetism, ferromagnetic order already exists in the core of
the Heusler layers, whereas at the interfaces spin glass order
still prevails. The coupling between the two types of magnetic
order will induce a small ferromagnetic magnetization close to the
interfaces, however, with some canting of the local magnetization
direction because of the coupling to the coexisting random spin
glass type of order. These canted interface moments might be
identified as the ``spin active zone'' needed for the conversion
of singlet pairs into triplet pairs \cite{EschrigTriplet}. Thus a
picture emerges which qualitatively resembles the three layer
ferromagnetic system in the theoretical work of
Ref. \cite{HouzetTriplet}, where triplet pairing is also found in a
narrow thickness range of the spin active interfaces only.

In summary, we have shown that making use of the unique magnetic
properties of \Cu-Heusler layers we can create a magnetization
profile inside the Heusler layer which can very effectively
generate triplet superconductivity. Aside from the
intriguing problem of the detailed microscopic origin of this
process, from a practical point of view we want to stress that the
preparation of Heusler films is perfectly compatible with
established preparation routes for Josephson junctions, unlike
other possible barrier materials like CrO$_{2}$ or Ho. Thus
Heusler layers open a new route to routinely prepare Josephson
junctions with nearly pure triplet supercurrents for applications
and basic research issues.
\vspace{-0.5cm}
\begin{acknowledgements}
 The authors thank the DFG for funding this work within the SFB 491 and within the DFG project WE 4359/1-1.
 Additionally we thank A. F. Volkov, M. Fistoul and K. B. Efetov for valuable discussions.
\end{acknowledgements}

\bibliographystyle{apsprl}

\end{document}